\documentclass{aa}

\usepackage{graphicx}
\usepackage{txfonts}
\usepackage{hyperref}
\newcommand{\tabhead}[1]{\textbf{#1}}
\emergencystretch=\maxdimen
\begin{document}

   \title{High-precision broadband linear polarimetry of early-type binaries \\ III. AO~Cassiopeiae revisited\thanks{The polarization data for AO~Cas are only available in electronic form at the CDS via anonymous ftp to \url{cdsarc.cds.unistra.fr} (\url{130.79.128.5}) or via \url{https://cdsarc.cds.unistra.fr/cgi-bin/qcat?J/A+A/}}}
        

    \author{Yasir~Abdul Qadir\inst{1}
    \and  Andrei V. Berdyugin\inst{1}
    \and Vilppu Piirola\inst{1}\and\\
    Takeshi Sakanoi\inst{2}
    \and Masato Kagitani\inst{2}}

    \institute{Department of Physics and Astronomy, FI-20014 University of Turku, Finland \\
    \email{yasir.abdulqadir@utu.fi}
    \and Graduate School of Sciences, Tohoku University, Aoba-ku, 980-8578 Sendai, Japan 
    }

   \date{Received 12 November 2022 / Accepted 17 January 2023}


    \abstract{}{The fact that the O-type close binary star system AO~Cassiopeiae exhibits variable phase-locked linear polarization has been known since the mid-1970s. In this work, we re-observe the polarization arising from this system more than 50 years later to better estimate the interstellar polarization and to independently derive the orbital parameters, such as inclination, $i$, orientation, $\Omega$, and the direction of the rotation for the inner orbit from the phase-folded polarization curves of the Stokes $q$ and $u$ parameters.}
    {The Dipol-2 polarimeter was used to obtain linear polarization measurements of AO~Cassiopeiae in the $B$, $V$, and $R$ passbands with the T60 remotely controlled telescope at an unprecedented accuracy level of $\sim$0.003\%. We have obtained the first proper quantification of the interstellar polarization in the direction heading towards AO~Cas by observing the polarization of three neighboring field stars. We employed a Lomb-Scargle algorithm and detected a clear periodic signal for the orbital period of AO~Cas. The standard analytical method based on a two-harmonics Fourier fit was used to obtain the inclination and orientation of the binary orbit.}
    {Our polarimetric data exhibited an unambiguous periodic signal at 1.76 days, thus confirming the orbital period of the binary system of 3.52 days. Most of the observed polarization is of interstellar origin. The de-biased values of the orbital inclination are $i = 63^{\circ} +2^{\circ}/ -3^{\circ}$ and orientation of $\Omega = 29^{\circ} (209^{\circ}) \pm 8^{\circ}$. The direction of the binary system rotation on the plane of the sky is clockwise.}{}

        
        \maketitle
        
        \section{Introduction}\label{sec:intro}
        AO~Cassiopeiae (AO~Cas) is an early-type binary system that consists of a larger, but less massive O9-III primary and late O-type main sequence secondary, with an orbital period of about 3.5 days and a mass ratio of $q = M_{s}/M_{p}= 1.47\pm 0.08$ \citep{GW}. The AO~Cas light curve exhibits only grazing eclipses and most of the light variations are due to geometrical distortion of the primary component. Apparently, the primary star fills its Roche lobe \citep{HH}, or is close to filling it \citep{GW}, and the system is likely to be engaged in a slow mass-exchange stage \citep{PLVC}. There is a clear spectroscopic evidence for the presence of high-density gas in the system and the behavior of the $H_{\rm \alpha}$ profile indicates the existence of the ionized bow shock area where the stellar winds collide \citep{GW}. 
        
        Because of its brightness, AO~Cas has been an early target for polarimetric observations. The first attempt to observe polarization from this binary system was made by Shakhovskoj (1964), who came to the conclusion that AO~Cas did not exhibit any variable polarization. This was due to the fact that the observational errors were rather large $\sim$0.09\% and no filters were used to conduct those observations. 
        
        The first successful variable polarimetric signals synchronous with the orbital phase from AO~Cas were detected and reported by \citet{PK1}, \citet{PK2}, and \citet{PK3}. It was then proposed by \citet{PFR} that the variable phase-locked polarization arises from scattering material within a large circumstellar envelope. The observed polarization is altered at times by events related to the mass exchange process within the binary system. It was further suggested that this large circumstellar envelope contains most of the scattering material, which is highly ionized. A suggestion was made to continue observing the system in order to track any changes that could occur in the observed polarization of the system \citep{PFR}. 
        
        A substantially better  polarization data set for AO~Cas, in terms of in quality and accuracy,  has been obtained by \citet{RK}. Their data analysis led to the conclusion that polarization arises from the light scattering on the gaseous stream, with the main contribution to polarization coming from the relatively small region located on the advancing side of the secondary star \citet{RK}. Later, these data were re-used by \citet{RK78}, \citet{BME}, \citet{SAB} and \citet{AS82} for estimations of the orbital inclination as well as in discussions of the accuracy of the orbit inclinations derived for the binary systems from polarization data. No definite conclusions on the exact location of the light-scattering region have been made and the apparent controversy between two different interpretations remains unresolved.

        Since the 1970s, AO~Cas system has never been observed polarimetrically. Thus, the question about the apparent seasonal variations of intrinsic polarization raised by \citep{PFR} remained  unaddressed. There has been a level of uncertainty around the amount of interstellar (IS) component in the observed polarization of AO~Cas because two estimates of this polarization, obtained by \citet{PFR} and \citet{BME}, differ dramatically. Therefore, we have decided to   this binary system  to study the polarization variability in more detail, determine the amount of interstellar polarization with a good level of confidence, and compare new polarization measurements with the previous ones. In this paper, describe the results of our new high-precision, three-band $BVR$ polarimetry of AO~Cas. 
        
   
        \section{Polarimetric observations}\label{sec:observations}
        
        \subsection{DiPol-2}\label{sec:DiPol}
        We observed AO~Cas for a total of 39 nights, from October 26, 2020 until February 01, 2021, with the DiPol-2 polarimeter \citep{PIRLA14}, attached to the remotely controlled T60 telescope at Haleakal\={a} Observatory, Hawaii. The observational log is given in Table \ref{table:log}. 
        
        DiPol-2 uses two dichroic beam-splitters that split the incident light beam into the $B$, $V$, and $R$ passbands, which are simultaneously recorded by three CCD cameras. The polarization modulator is a rotatable superachromatic $\lambda/2$ (\rm or $\lambda/4$) plate and the polarization analyzer is a plane-parallel calcite plate. A typical cycle of linear polarimetric measurement consists of 16 exposures at the orientation intervals of $22.5^\circ$ of the $\lambda/2$ plate. The calcite plate produces two orthogonally polarized stellar images that are recorded simultaneously to eliminate errors that may arise due to varying atmospheric transparency; thus, the contribution from the sky polarization is automatically canceled out. 
        
        For observations of the bright stars, such as AO~Cas, DiPol-2 employs the intentional defocusing technique, whereby the stellar image is spread over the large number of pixels. This allows us to collect up to $10^{7}$ photo-electrons per exposure and avoid pixel saturation. Generally, around 180 -- 320 images were taken every night with the exposure time of 3s. This corresponds to 45 -- 80 measurements of Stokes $q$ and $u$ per night. The skyflat images that are used for calibration purposes were regularly taken at twilight hours, either in the beginning of the observing night or at dawn. Once per night, the series of dark and bias images were also taken. 
        
        The telescope polarization was derived from observations of 20--25 zero-polarized standard stars. It was found to be in the range between 0.004\% -- 0.006\%. The accuracy of the determination of instrumental polarization is 0.0002\% -- 0.0003\%. Polarization values for AO~Cas have been corrected for instrumental polarization. For the calibration of polarization angle zero-point, polarization angles of highly polarized ($5\%$ -- $6\%$) standard stars HD~204827 and HD~25443 were used. Their polarization angles are given in Table \ref{table:hp}.
        
        The resulting accuracy in the case of AO~Cas is at the level of 0.002\% -- 0.003\% in the $B$ passband and 0.003\% -- 0.005\% in the $V$ and $R$ passbands. Thus, our accuracy is on an order of magnitude that is higher than the value $\sigma_{\rm p} \sim$ 0.025\% that was previously obtained by \citet{RK}.

     \begin{table}
        \caption{Average polarization angles ($\theta$) of highly polarized stars.} 
        \label{table:hp}
        \centering
        \renewcommand{\arraystretch}{1.0} 
        \scalebox{1.0}{
        \begin{tabular}[c]{ l c c c} 
                \hline\hline 
                Star & Passband & $\theta$~[deg] & References \\ \hline
                HD~204827  & $B$ & $57.79 \pm 0.02$ & [1] \\ 
                & $V$ & $58.33 \pm 0.02$ & [1] \\  
            & $R$ & $59.21 \pm 0.02$ & [1] \\
                HD 25443 & $B$ & $134.28 \pm 0.51$ & [2] \\
                & $V$ & $134.23 \pm 0.34$ & [2] \\
                & $R$ & $133.65 \pm 0.28$ & [2] \\
                \hline
        \end{tabular}}
        \tablebib{(1)~\citet{PRL21}; (2) \citet{SCHMDT}.}
   \end{table} 
        
        \begin{figure*}
                \centering
                \includegraphics[width=1\textwidth]{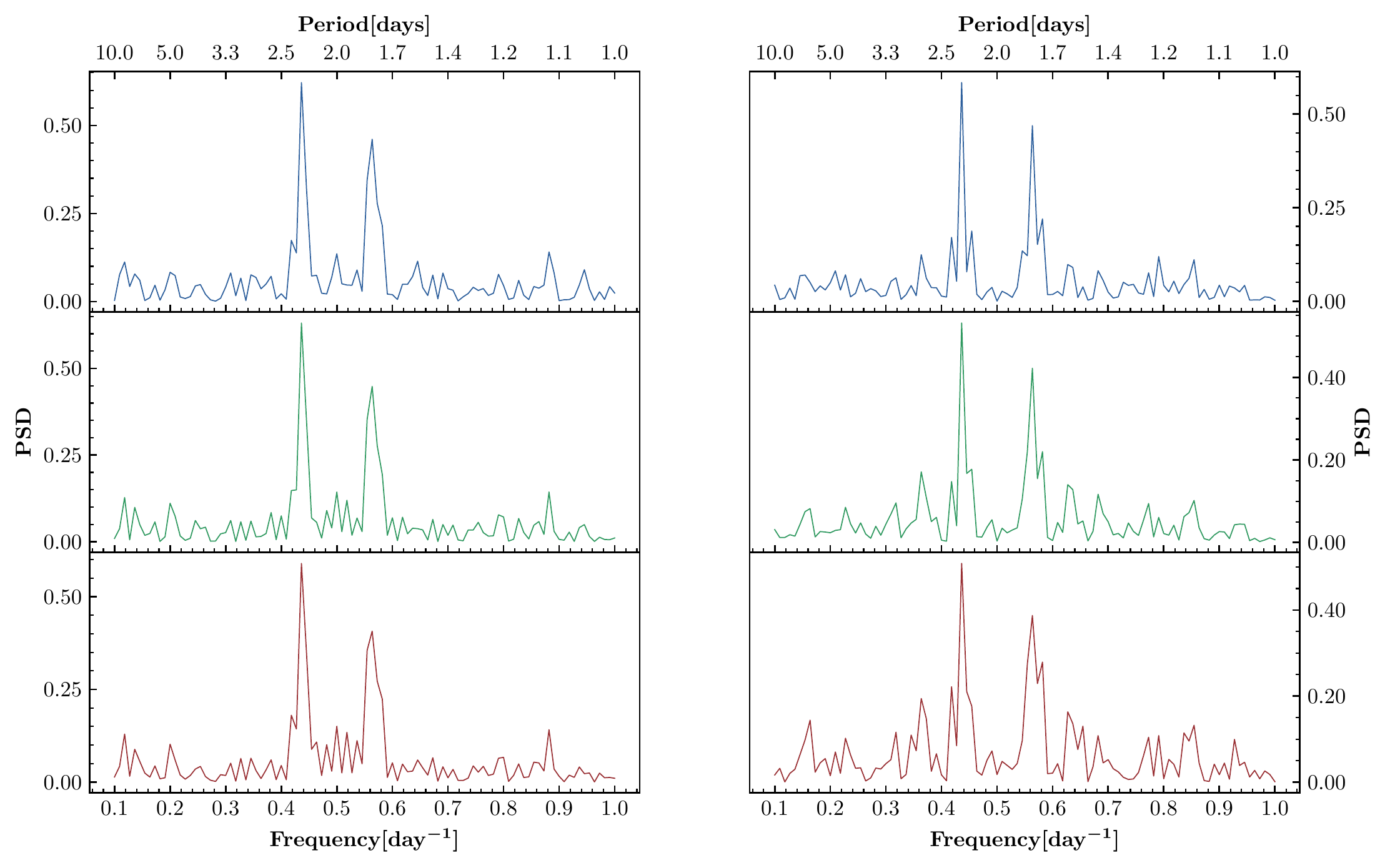}
                \caption{Lomb-Scargle periodograms for Stokes $q$ (left column) and $u$ (right column) of AO~Cas in $B$, $V$, and $R$ passbands (top, middle and bottom panels respectively).} 
                \label{fig:ls}
        \end{figure*}

        \subsection{Data reduction}\label{sec:reduction}
        The subtraction of bias and dark frames and the applicaton of flat-fielding in order to eliminate any spurious effects arising from these imperfections was performed via automated Visual Basic scripts that are capable of processing several hundreds of polarimetry CCD images simultaneously. A special algorithm has been developed to pre-align images taken in long series of polarimetric sequences to remove possible effect of the image drift. The scripts, executed with the MaxIM DL Pro\footnote{\url{https://diffractionlimited.com/downloads/GettingStarted.pdf}}, perform the calibration and create  2 x 2 binned sub-frames for every taken raw image.      
           
        Normalized Stokes parameters $q$ and $u$ are computed from the flux intensity ratios of the orthogonally polarized stellar images $Q_{\rm i} = I_{\rm e}(i)/I_{\rm o}(i)$ obtained for each orientation of wave plate, $i$ = $0.0^{\circ}$, $22.5^{\circ}$, $45.0^{\circ}$, $67.5^{\circ}$, as: 
        
        \begin{equation}
                \begin{split}
                        Q_{\rm m} = Q_{0.0} + Q_{22.5} + Q_{45.0} + Q_{67.5}, \\
                        q  = (Q_{0.0} - Q_{45.0})/Q_{\rm m}, \\
                        u = (Q_{22.5} - Q_{67.5})/Q_{\rm m}.
                        \label{Intensity ratios, Stokes q, Stokes u}
                \end{split}
        \end{equation}

        Because AO Cas is a bright target, intentional defocusing was used (see \citet{PIRLA14} and \citet{PRL21}). Typical software aperture size for the defocused stellar image was chosen as $\sim$7 -- 10~arcsec. This allows us to collect up to $10^7$~ADUs per image, avoiding image saturation while keeping the o and e-images well separated (and not overlaid). The distance between the centers of o and e-images in the focal plane is $\sim$22~arcsec. The intensities $I_{\rm e}(i)$ and $I_{\rm o}(i)$ were determined from the series of calibrated polarimetry images with the aperture photometry method implemented as a standard feature in the MaxIM DL Pro software. A simple \textsc{Fortran} code was used to compute the Stokes parameters $q$ and $u$ from each set of four images. This code also computes weighted average values of polarization. It gives lower weight to individual measurements of Stokes $q$ and $u$ that deviate from the mean value by more than expected from Gaussian noise distribution. All the measurement points that were within $2\sigma$ were given an equal weight and any points that deviated more $2\sigma$ were given a lower weight that was proportional to the inverse square of the estimated error. Any points beyond $3\sigma$ were rejected. Using Stokes parameters $q$ and $u$, the code computes the values of polarization $P$ and polarization angle $\theta$.

    \begin{table}
        \caption{FAP for AO~Cas Periodograms.} 
        \label{table:fap}
        \centering
        \renewcommand{\arraystretch}{1.0} 
        \begin{tabular}[c]{ l c c c} 
                \hline\hline 
                & Passband & Period~[days] & FAP \\ \hline
                Stokes $q$: & $B$ & 1.76 & $1.84 \times 10^{-10}$ \\ 
                && 2.31 & $1.75 \times 10^{-9}$ \\
                & $V$ & 1.76 & $5.7 \times 10^{-11}$ \\  
                && 2.31 & $2.88 \times 10^{-10}$ \\  
        & $R$ & 1.76 & $1.67 \times 10^{-10}$ \\
        && 2.31 & $1.16 \times 10^{-9}$ \\
                Stokes $u$: & $B$ & 1.76 & $9.5 \times 10^{-11}$ \\  
                && 2.31 & $8.0 \times 10^{-9}$ \\
                & $V$ & 1.76 & $1.28 \times 10^{-8}$ \\ 
                && 2.31 & $1.01 \times 10^{-7}$ \\
                & $R$ & 1.76 & $6.95 \times 10^{-8}$ \\
                && 2.31 & $2.78 \times 10^{-6}$ \\
                \hline
        \end{tabular}
   \end{table}

        \section{Data analysis}\label{sec:analysis}
        \subsection{Period search}\label{sec:prdsch}
        After re-observing AO~Cas polarimetrically after a time span of more than 50 years, we decided to perform a rigorous period search utilizing our new high-precision polarization data. We employed a Lomb-Scargle algorithm \citep{LMB76, SCRGL82} from \texttt{astropy.timeseries}\footnote{\url{https://docs.astropy.org/en/stable/timeseries/ lombscargle.html}} \citep{PRCW18} in \textsc{Python} to determine any periodic signals which may present in polarimetric data. The advantage in using Lomb-Scargle periodogram is that the algorithm utilizes a least-squares method to fit a sinusoidal on unevenly sampled data (which it is often the case with respect to astronomical data).

        Due to the orbital motion, the degree of intrinsic polarization from the scattered light in a binary system usually reaches maximum when the scattering angle is near $90^\circ$ or $270^\circ$ and minimum when it is near $0^\circ$ or $180^\circ$. Therefore, Lomb-Scargle periodograms applied to polarization data is expected to detect only half of the orbital period for systems such as AO~Cas. To visualize the periodic search in our data, we plotted Lomb-Scargle periodograms for both Stokes  $q$ and $u$ in $B$, $V$, and $R$ passbands (Figure \ref{fig:ls}). All these periodograms show two very prominent peaks: 1) at 1.76 days, namely, half of the known orbital period of AO~Cas; and 2) the nearby peak at 2.31 days. 
        
        \begin{figure*}
    \centering
    \includegraphics[width=1.0\textwidth]{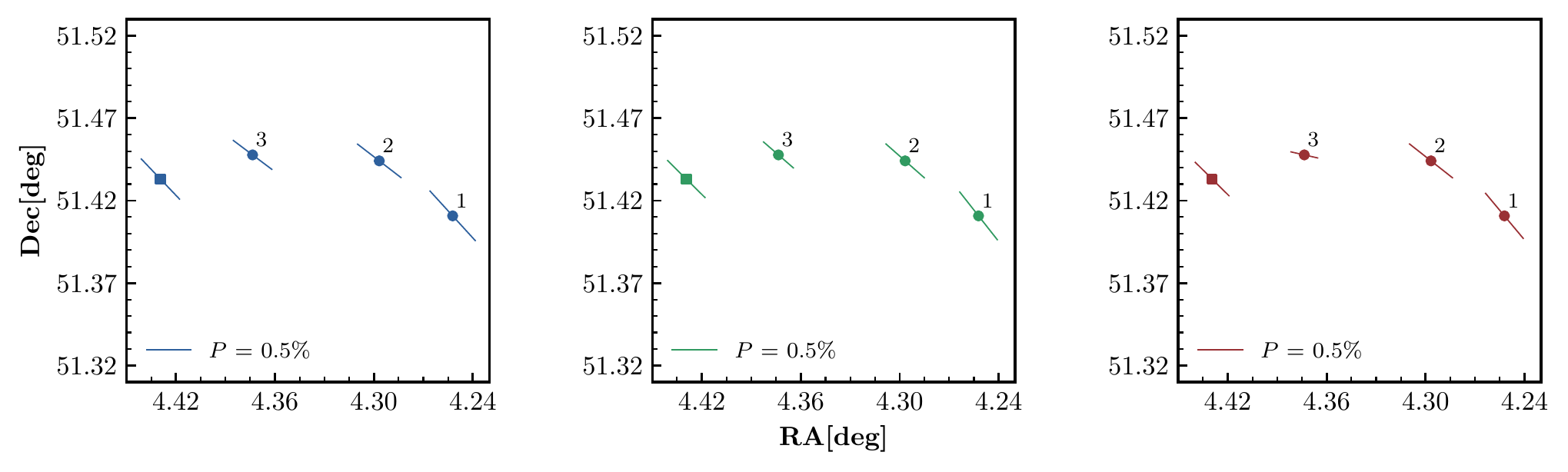}
    \caption{Polarization map of AO Cas (square) and field stars (circles) in $B$ (left panel), $V$ (middle panel), and $R$ (right panel) passbands. The length of the bars corresponds to the degree of observed polarization $P$, and the direction corresponds to the polarization angle (measured from the north to the east). The field stars are numbered the same as in Table \ref{table:is}.} 
    \label{fig:ref}
    \end{figure*}

    \begin{figure*}
                \centering
                \includegraphics[width=1.0\textwidth]{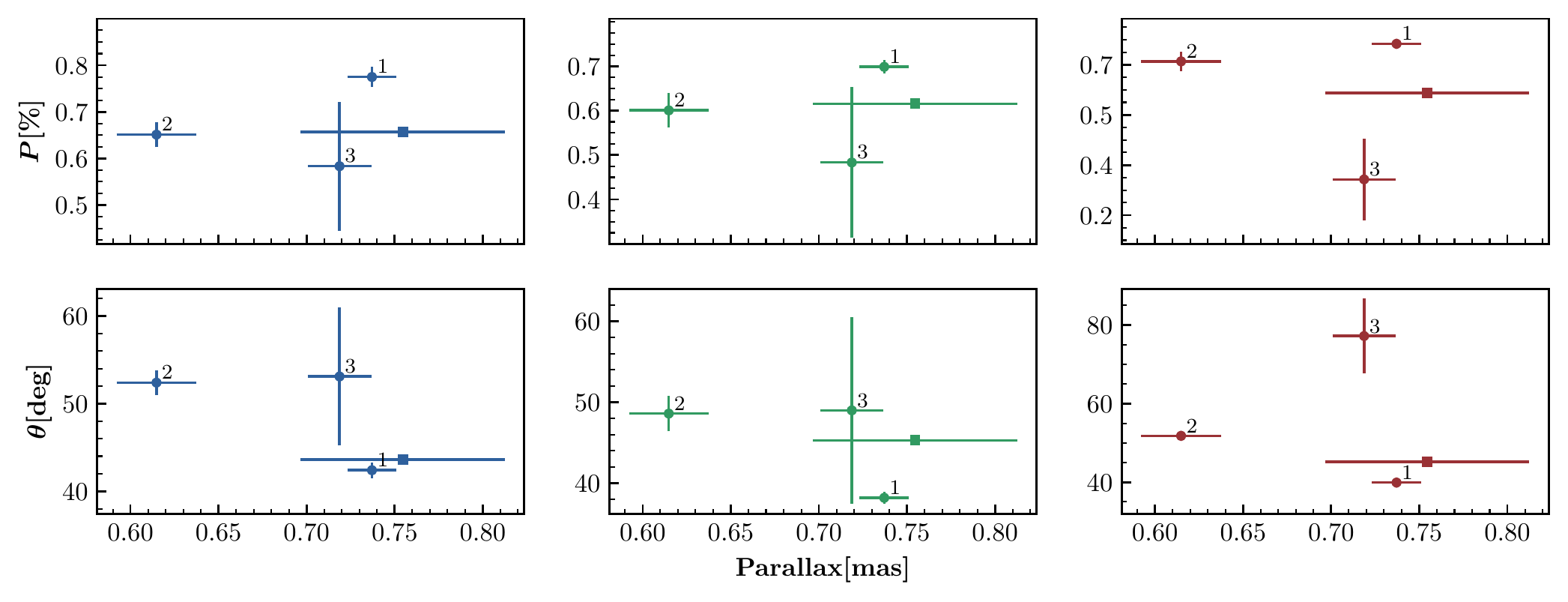}
                \caption{Dependence of the observed degree of polarization $P$ (top panels) and polarization angle $\theta$ (bottom panels) in $B$ (left panels), $V$ (middle panels), and $R$ (right panels) passbands on parallax of AO~Cas (square) and field stars (circles) with reference numbers that are same as in Table \ref{table:is}. The error bars correspond to $\pm\sigma$ errors.} 
                \label{fig:refs}
        \end{figure*}
        
        \begin{table*}[htp!]
                \centering
                \caption{Identifiers, coordinates, parallaxes, Stokes parameters, polarization degrees and angles, and number of exposures for each field star.}
                \label{table:is}
                \centering
                \renewcommand{\arraystretch}{1.0} 
                \scalebox{1.0}
                {
                        \begin{tabular}[c]{ l c c c c c c c c c c }
                                \hline\hline 
                                Identifier & Coordinates  & Parallax & Passband & Stokes $q$ & Stokes $u$ & $P$ & error\tablefootmark{~a} & $\theta$ & $N_{\rm exp}$ \\
                                
                                [Gaia DR3] & [J2000d] & [mas] & & [\%] & [\%] & [\%] & [\%] & [deg] \\
                                \hline
                                
                395017526722293248 & 4.252220267, & 0.7371 & $B$ & 0.073 & 0.809 & 0.812 & 0.026 & $42.4\pm0.9$ & 320 \\
                (Ref 1)& 51.41073703 & $\pm0.014$ & $V$ & 0.174 & 0.715 & 0.736 & 0.018 & $38.2\pm0.7$ & 320 \\
                &&& $R$ & 0.125 & 0.699 & 0.710 & 0.009 & $39.9\pm0.4$ & 320 \\
                
                395017801600190336 & 4.29671658, & 0.6148 & $B$ & -0.169 & 0.642 & 0.664 & 0.032 & $52.4\pm1.4$ & 128 \\
                (Ref 2) & 51.44408771 & $\pm0.023$ & $V$ & -0.078 &  0.613 & 0.618 & 0.047 & $48.6\pm2.2$ & 128 \\
                &&& $R$ & -0.156 & 0.643 & 0.661 & 0.026 & $51.8\pm1.1$ & 128 \\
                
                394976810432321792 & 4.37352903, & 0.7187 & $B$ & -0.162 & 0.560 & 0.583 & 0.166 & $53.9\pm7.9$ & 64 \\
                (Ref 3) & 51.44770379 & $\pm0.018$ & $V$ & -0.066 & 0.473 & 0.478 & 0.203 & $49.0\pm11.5$ & 64 \\
                &&& $R$ & -0.300 & 0.143 & 0.332 & 0.114 & $77.2\pm9.5$ & 64 \\
                                \hline  
                        \end{tabular}
                } 
                \tablefoot{
        \tablefoottext{a}{The error value is same for Stokes $q$, Stokes $u$, and $P$ in each
        corresponding row.}}
        \end{table*}

    \begin{table*}
                \centering
                \caption{ Average observed  $P_{\rm obs}$, $\theta_{\rm obs}$, average interstellar $P_{\rm is}$, $\theta_{\rm is}$, and average intrinsic $P_{\rm int}$, $\theta_{\rm int}$ of AO~Cas.}
                \label{table:refs}
                \centering
                \renewcommand{\arraystretch}{1.0} 
                \scalebox{1.0}
                {
                        \begin{tabular}[c]{ l c c c c c c }
                                \hline\hline 
                                Passband & $P_{\rm obs}$ & $\theta_{\rm obs}$ & $P_{\rm is}$ & $\theta_{\rm is}$ & $P_{\rm int}$ & $\theta_{\rm int}$ \\
                                & [\%] & [deg] & [\%] & [deg] & [\%] & [deg] \\
                                \hline
                $B$ & $0.671\pm0.003$ & $43.6\pm0.1$ & $0.741\pm0.021$ & $45.9\pm0.8$ & $0.091\pm0.021$ & $154.3\pm0.1$ \\
                
                $V$ & $0.636\pm0.002$ & $45.3\pm0.1$ & $0.714\pm0.017$ & $39.3\pm0.7$ & $0.161\pm0.017$ & $101.6\pm0.1$ \\
                
                $R$ & $0.573\pm0.003$ & $45.2\pm0.1$ & $0.698\pm0.010$ & $41.2\pm0.4$ & 
                $0.153\pm0.010$ & $115.4\pm0.1$ \\
                                \hline  
                        \end{tabular}
                } 
        \end{table*}

    We believe that the peak seen at 2.31 days is nothing but an alias peak of the main peak at 1.76 days. The alias peaks can be produced by periodicity in the timing of observations coupled with the periodic nature of the source that is being observed. If that period is not less than half of the sampling frequency, that is, the Nyquist frequency ($f_{\rm ny}$), then there is a side effect that produces two waves that differ by 1/$f_{\rm ny}$. With the (half) orbital period for AO~Cas, $\sim$1.76 days, which corresponds to the frequency of 0.57; this is more than half of the sampling frequency of $\sim$0.4, as we used data from 39 nights that were observed over a period of about 100 days. Therefore, an alias peak can be expected at 1/(1-1/1.76) = 2.31 days, which is exactly where the periodograms show the alias peaks (cf. \citet{SHNON49}; \citet{VPLS18}). A similar phenomenon was observed by \citet{KSNKV18} when deducing superhump period of the black hole X-ray binary GX 339--4. 
        
        Furthermore, we calculated the false alarm probability (FAP) by using a bootstrap method \citep{SUVG} for peaks associated with the orbital period of AO~Cas as well as their alias peaks. These values are given in Table \ref{table:fap}, which shows that FAP for the peaks associated with the "polarimetric" orbital period is very low, but the same is true for alias peaks as well, since the algorithm cannot distinguish between the real peak and its alias peak. Since the orbital period of AO Cas is already known to be 3.5~days, we can conclude that the real peak is the one seen at 1.76~days. Therefore, it can be deduced that the period search algorithm applied to the new polarization data has clearly revealed the orbital period of the binary system. We have not detected any other periods in our polarization data of AO~Cas.

        \subsection{Interstellar polarization}\label{sec:isp}
        
        The observed polarization $P_{\rm obs}$ of almost every distant star contains not only the intrinsic polarization component, but also the interstellar (IS) polarization component, $P_{\rm is}$, which is due to interstellar dust. Since the observed average polarization of AO~Cas is rather high ($\sim$0.6\%) and the distance to the binary is large ($\sim$1300~pc \citep{GAIA}), one can expect significant contribution of the $P_{\rm is}$ component. The IS polarization for AO~Cas has been estimated by \citep{PFR} for the $B$ and $V$ passbands as $P_{\rm is} = 1.2 \pm 0,2\%$, $\theta_{\rm is} = 
        80^{\circ} \pm 5^{\circ}$. After subtraction of IS component, Pfeiffer got intrinsic polarization value for AO~Cas peaking at $P_{\rm int} \ge 1.5\%$, (cf. Figs. 2 and 3 from \citep{PFR}).  Later on, \citep{BME} re-evaluated IS polarization component for AO~Cas and got significantly smaller value of $P_{\rm is} = 0.5\%$ with the direction $\theta_{\rm is} = 55^{\circ}$.
        
            \begin{figure*}
                \centering
                \includegraphics[width=1\textwidth]{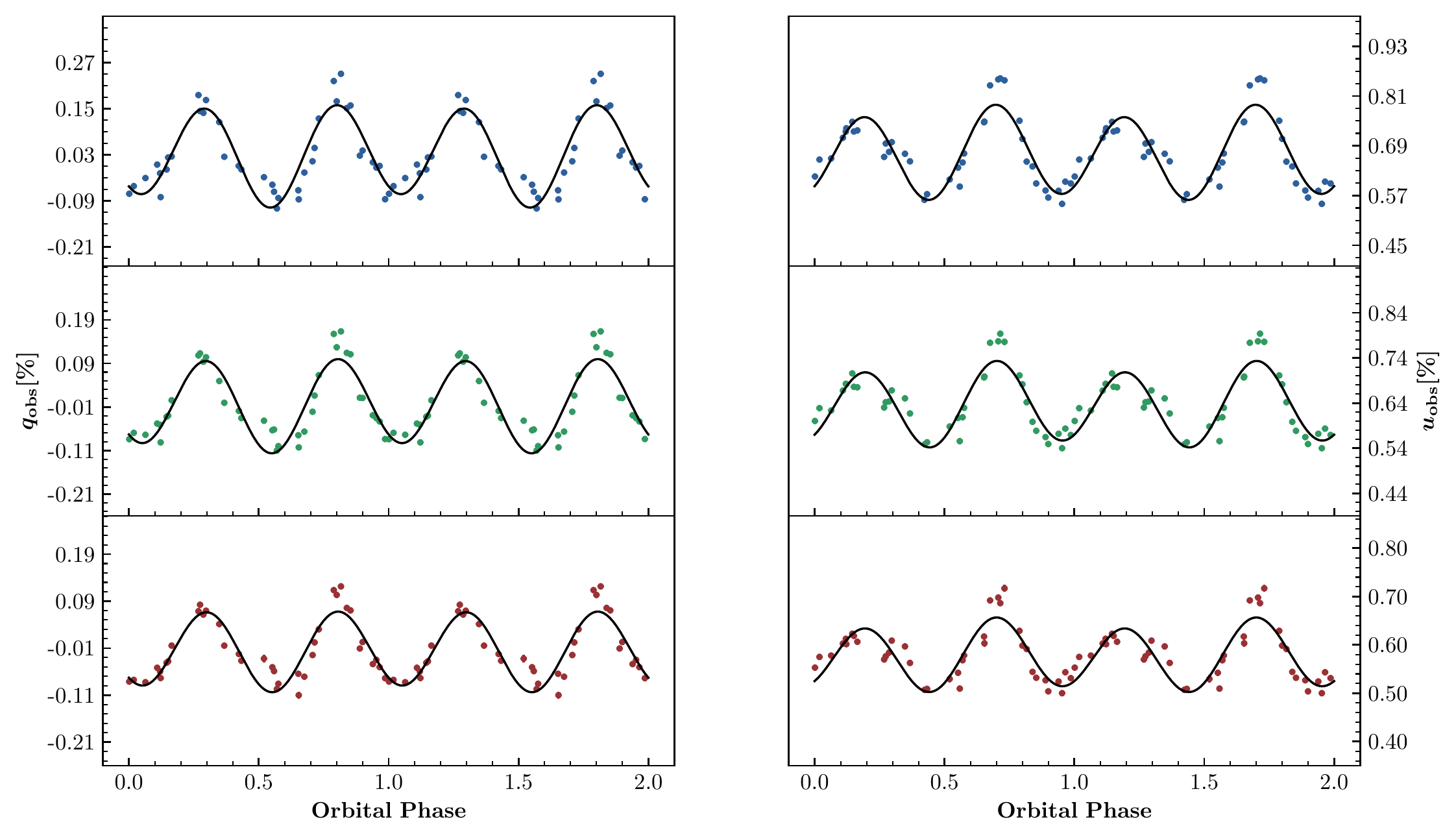}
                \caption{Variability of observed Stokes $q$ and $u$ parameters for AO~Cas in $B$, $V$, and $R$ passbands shown in the top, middle, and bottom panels, respectively, which are phase-folded at the orbital period of 3.52 days. Fourier fit curves (see Sect. \ref{sec:fcm}) are shown in solid lines and the best fit Fourier coefficients are given in Table \ref{table:fc}. For the majority of data points, the length of $\pm\sigma$ error bars is shorter than the size of the plotting symbols.} 
                \label{fig:fs}
        \end{figure*}
        
        \begin{table*}
                \centering
                \caption{Best-fit Fourier coefficients for Stokes $q$ and $u$.}
                \label{table:fc}
                \centering
            \renewcommand{\arraystretch}{1.0} 
                \scalebox{1.0}
                {
                        \begin{tabular}[c]{ l c c c c c c c c c c }
                                \hline\hline 
                                \tabhead{\shortstack{\rm Passband }} & \tabhead{\shortstack{\rm $q_0$}}  & \tabhead{\shortstack{\rm $q_1$}} 
                                & \tabhead{\shortstack{\rm $q_2$}} & \tabhead{\shortstack{\rm $q_3$}} & \tabhead{\shortstack{\rm $q_4$}}
                                & \tabhead{\shortstack{\rm $u_0$}} & \tabhead{\shortstack{\rm $u_1$}} & \tabhead{\shortstack{\rm $u_2$}}
                                & \tabhead{\shortstack{\rm $u_3$}} & \tabhead{\shortstack{\rm $u_4$}} \\                              
                \hline
                                $B$ & 0.0304 & 0.0181 & 0.0007 & -0.1026 & -0.0664
                                & 0.6694 & 0.0017 & -0.0165 & -0.0796 & 0.0662 \\
                                & $\pm$0.0067 & $\pm$0.0095 & $\pm$0.0094 & $\pm$0.0098 & $\pm$0.0090 
                                & $\pm$0.0055 & $\pm$0.0079 & $\pm$0.0078 & $\pm$0.0081 & $\pm$0.0075 \\
                                
                                $V$ & -0.0077 & 0.0118 & 0.0015 & -0.0814 & -0.0600
                                & 0.6387 & 0.0031 & -0.0144 & -0.0683 & 0.0515 \\
                                & $\pm$0.0053 & $\pm$0.0077 & $\pm$0.0075 & $\pm$0.0079 & $\pm$0.0073 
                                & $\pm$0.0049 & $\pm$0.0070 & $\pm$0.0069 & $\pm$0.0073 & $\pm$0.0067 \\                          
                                
                                $R$ & -0.0155 & 0.0069 & 0.0016 & -0.0650 & -0.0497
                                & 0.5736 & 0.0020 & -0.0127 & -0.0539 & 0.0420 \\
                                & $\pm$0.0042 & $\pm$0.0061 & $\pm$0.0058 & $\pm$0.0062 & $\pm$0.0056 
                                & $\pm$0.0044 & $\pm$0.0063 & $\pm$0.0060 & $\pm$0.0064 & $\pm$0.0058 \\                          
                                \hline  
                        \end{tabular}} 
        \end{table*}

        In order to obtain more reliable estimates for the $P_{\rm is}$ and $\theta_{\rm is}$, we decided to conduct polarimetric observations of field stars in the close proximity of AO~Cas. We used the GAIA database \footnote{\url{https://gea.esac.esa.int/archive/}} to search for stars within the angular distances of $\leq7^{'}$ from AO~Cas and parallaxes between 0.6 -- 0.8~mas, as the parallax of AO~Cas itself is $0.76 \pm 0.06$~mas \citep{GAIA}. Three such stars have been found and observed. 
        
        In Figure \ref{fig:ref}, we have plotted the polarization maps that show the distribution of IS polarization vectors around AO Cas in $B$, $V$, and $R$ passbands. Table \ref{table:is} gives their Gaia identifiers, coordinates, parallaxes, Stokes parameters, polarization degrees and angles, and the total number of exposures. Figure \ref{fig:refs} shows the dependence of field stars polarizations and polarization angles on the parallax. Using these data, we have calculated weighted mean values for the IS polarization. In Table \ref{table:refs}, we are showing average observed polarization degree $P_{\rm obs}$ and angle $\theta_{\rm obs}$, the interstellar polarization degree $P_{\rm is}$ and angle $\theta_{\rm is}$, and the average intrinsic polarization degree, $P_{\rm int}$ and angle $\theta_{\rm int}$, together with their errors in each passband for AO~Cas. We used the average of all nightly measurements for both AO~Cas and field stars where we observed the data for more than one night.
        
        As is seen from Table \ref{table:refs}, the IS polarization degree derived by us, $P_{\rm is} \sim 0.7\%$ is significantly lower than that derived by Pfeiffer, but larger than derived by \citep{BME}. The new IS polarization angle, $\theta_{\rm is} \sim 40^{\circ}$, is also different from both of the previously determined values. Given the new reliable estimates of the IS polarization parameters, it becomes obvious that most of the observed polarization in AO~Cas is due to interstellar dust. The value of intrinsic polarization in this binary is rather small: $P_{\rm int} \simeq 0.1 - 0.2\%$. It appears that previous estimates for $P_{\rm int} > 1.0\%$ made by \citep{PFR} were based on wrong adopted parameters of the IS polarization component. Thus, the presence of a large circumstellar envelope, which would otherwise need to exist to produce such a large intrinsic polarization, can be ruled out.      
        
        \subsection{Polarization variability}\label{sec:fcm}
        The common method of analysis of the phase-locked binary polarization is fitting the phase curves of the Stokes parameters $q$ and $u$ with Fourier series expansion. This method is also known as the "BME" approach, \citep{BME}. For a circular orbit with co-rotating light scattering envelope, this fit includes only zeroth, first and second harmonics terms:
        
        \begin{equation}
                \begin{split}
                q =  q_0 + q_1 \cos \lambda + q_2 \sin \lambda + q_3 \cos 2\lambda + q_4 \sin 2\lambda, \\
                u =  u_0 + u_1 \cos \lambda + q_2 \sin \lambda + q_3 \cos 2\lambda + q_4 \sin 2\lambda,
                \end{split}
        \label{Fourier Series}
        \end{equation}
        
         where, $\lambda =  2 \pi \phi$ and $\phi$ is a phase of the orbital period. We used \texttt{curve$\_$fit} function of \texttt{scipy.optimize}\footnote{\url{https://docs.scipy.org/doc/scipy/reference/optimize.html}} library in \textsc{Python} to obtain best fit parameters together with their errors. They are shown in Table \ref{table:fc}. After deriving Fourier coefficients from the best fits to each passband, we plotted the fitted curves over the observational data points in Figure \ref{fig:fs}. The ephemeris which has been used is: $P_{\rm orb}~[\rm days] = 3.52348$, $T_0~[\rm MJD] =2445294.97083$\footnote{\url{https://www.aavso.org/ bob-nelsons-o-c-files.}}. Phase 0.0 corresponds to the primary minimum. 
         
         It is possible to derive the orbital inclination $i$ using  the values of the first ($q_{1,2}$, $u_{1,2}$) and second ($q_{3,4}$, $u_{3,4}$) harmonics terms of Fourier series with the following formulas \citep{DRSN}:
        
        \begin{equation}
        \begin{split}
        \begin{aligned}
        & \left(\frac{1 - \cos i}{1 + \cos i}\right)^4 =  \frac{(u_1 + q_2)^2 + (u_2 - q_1)^2}{(u_2 + q_1)^2 + (u_1 - q_2)^2}, \\
        & \rm or \\
        & \left(\frac{1 - \cos i}{1 + \cos i}\right)^4 =  \frac{(u_3 + q_4)^2 + (u_4 - q_3)^2}{(u_4 + q_3)^2 + (u_3 - q_4)^2}.                
        \label{Inclination}
        \end{aligned}
        \end{split}
        \end{equation}
        
        In the case of a circular orbit and distribution of the light scattering material symmetric to the orbital plane, the first-harmonics terms are negligibly small and only the second-harmonics terms can reliably be used for the determination of the orbital parameters. In addition to the orbital inclination, the orientation of the orbit on the sky (longitude of ascending node  $\Omega$) can be found from \citep{DRSN} as follows: 
        
        \begin{equation}
        \tan \Omega = \frac{A+B}{C+D} = \frac{C-D}{A-B},
        \label{Inclination}
        \end{equation}

        where,
        
        \begin{equation}
                \begin{split}
                A = \frac{u_4 - q_3}{(1 - \cos i)^2}, \:\:\:\:\:\:\:\:\:\: B = \frac{u_4 + q_3}{(1 + \cos i)^2}, \\
                C = \frac{q_4 - u_3}{(1 + \cos i)^2}, \:\:\:\:\:\:\:\:\:\: D = \frac{u_3 + q_4}{(1 - \cos i)^2}.
                \end{split}
        \label{alphabets}
        \end{equation}

        Moreover, $A_{\rm q}$, and $A_{\rm u}$ are two parameters which define the ratio of the amplitudes of second to first harmonics for Stokes $q$ and $u,$ respectively. For a circular orbit, these parameters are effective measures of the degree of symmetry and the concentration scattering material towards the orbital plane. Their values were obtained from the following formulae:
        
        \begin{equation}
                \begin{split}
                        A_{\rm q} = \sqrt{\frac{q_3^2 + q_4^2}{q_1^2 + q_2^2}} , \:\:\:\:\: A_{\rm u} = \sqrt{\frac{u_3^2 + u_4^2}{u_1^2 + u_2^2}}. 
                \end{split}
                \label{Ratios}
        \end{equation}
        
     With Eqs. 3-6, we have derived the values of $i$, $\Omega$, $A_{\rm q}$, and $A_{\rm u}$ for the $B$, $V$, and $R$ passbands, given in Table \ref{table:orbpar}. As is seen from this table, second harmonics variations are clearly dominating in the polarization variability of AO~Cas. 
     
\begin{figure}
    \centering
    \includegraphics[width=0.5\textwidth]{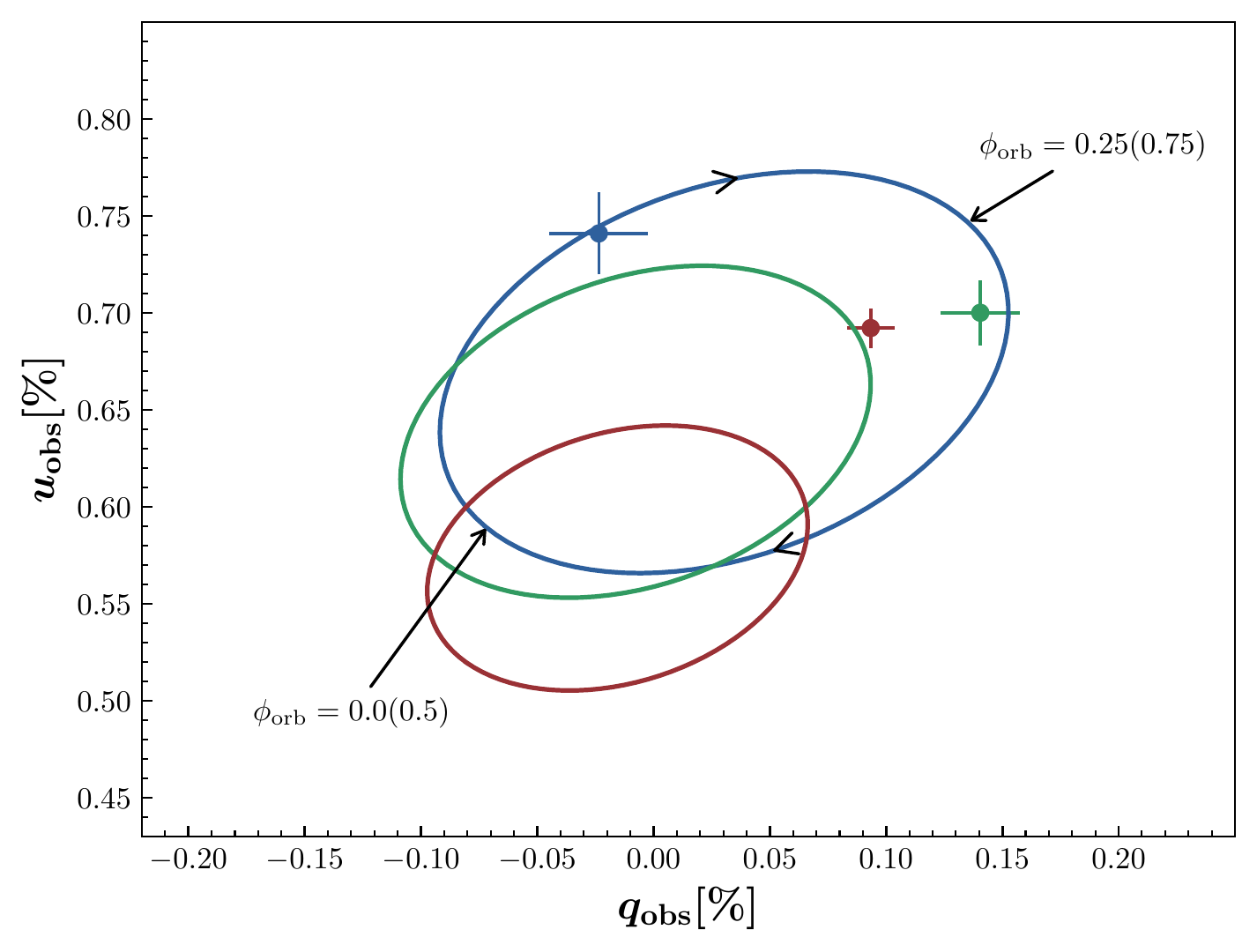}
    \caption{Variability of observed polarization for AO~Cas plotted on the Stokes ($q$, $u$) plane, represented by the ellipses of second harmonics of Fourier fit. The clockwise direction and phases of the orbital period are depicted on $B$ passband ellipse. The angle between the major axis and the $q$-axis gives the orientation $\Omega$. Average interstellar Stokes $q$ and $u$ parameters are depicted by circles with $\pm\sigma$ error bars. Blue, green, and red colors represent $B$, $V$, and $R$ passbands.} 
    \label{fig:elps}
    \end{figure}

     Figure \ref{fig:elps} shows the ellipses of the second harmonics on the ($q$, $u$) plane, with the derived Stokes parameters of the IS polarization for the $B$, $V$, and $R$ passbands. The eccentricity of the ellipses is related to the inclination of the orbit, $i$, and the orientation of their major semi-axes with respect to $q$-axis defines the orientation of the orbit, $\Omega$. The direction of circumvention corresponds to the direction of the orbital motion in the binary system on the plane of the sky. In the case of AO~Cas, this direction is clockwise.
     
     In addition to the strong second harmonics, there is a small, but non-zero first harmonics in the variability of Stokes parameter, u, at the significance level of $\sim$2$\sigma$. We note, that the presence of the first harmonics is also apparent in the data obtained by \citet{RK}. Thus, the slight asymmetry of light scattering material with respect to the orbital plane in the AO~Cas appears to be real. 
     
     Comparison of the $B$ passband polarization variability curve of the Stokes parameters shown in our Figure \ref{fig:fs}, with that published by \citet{RK}\footnote{Note: the variability "curves" plotted by Rudy \& Kemp (1976) for their measured Stokes parameters are hand-drawn. }, reveals their remarkable similarity. Even the higher secondary peaks in Stokes $q$ and $u$ at the phase 0.75 are visible on the new and old plots, which are separated in time by half a century. In contrast to \citep{PFR}, our polarization phase curves do not show evidence for long-term effects, such as seasonal changes. Our new results are strongly suggesting that general distribution of light scattering material in the AO~Cas system remains stable over many decades. Therefore, the conclusion made by \citep{PFR} about noticeable cycle-to-cycle variability is most likely due to the low accuracy of his polarization data.

    \begin{table}
        \centering
        \caption{Orbital parameters of AO~Cas in $B$, $V$, and $R$ passbands.} 
        \label{table:orbpar}
    \renewcommand{\arraystretch}{1.0} 
        \begin{tabular}[c]{l  c c c} 
                \hline\hline 
                Passband & $i$\tablefootmark{~a} & $\Omega$ & $A_{\rm q}/A_{\rm u}$ \\ \hline
                $B$ & $63.67^{\circ}$ & $ 30.39^{\circ} (210.39^{\circ})$ & 6.75/6.25 \\  
                $V$ & $63.22^{\circ}$ & $ 29.82^{\circ} (209.82^{\circ})$ & 8.63/10.11 \\  
                $R$ & $62.34^{\circ}$ & $ 27.16^{\circ} (207.16^{\circ})$& 10.94/9.29 \\  
                \hline
        \end{tabular}\\
    \tablefoot{
    \tablefoottext{a}{The values given are de-biased and are $2^\circ$ lower than biased values.}}
    \end{table} 
     
     As is well known, due to the noise in polarimetric data arising from measurement uncertainties, the value of $i$ derived from the Fourier fit is always biased towards higher values (\citet{ASB}; \citet{SAB}; \citet{WD}). The amount of bias also depends on the value of true inclination, the lower the true value of $i$, the higher is the bias. A similar bias may be induced due to stochastic noise caused by intrinsic non-periodic component in polarization variations \citep{MB}. Obviously, the deviations of the data from the fit can also be due to departure of the binary viewing geometry from the BME model assumptions, such as the absence of stellar eclipses and/or occultations of the light scattering regions occurring at certain range of the orbital phase.     
    
    As is seen from Table \ref{table:orbpar}, the values of $i$ and $\Omega$ derived for the three passbands are in good agreement with each other. There is an ambiguity on the value of $\Omega$ as $\Omega + 180^\circ$ is equally possible \citep{DRSN}. Because AO~Cas is a (partially) eclipsing binary, the true value of inclination is rather high, and that should mean a smaller bias. In order to account for such bias, we derived confidence intervals for $i$ and $\Omega$ defined by \citet{WD}. This method utilizes the special merit parameter $\gamma,$ given as follows: 
    
    \begin{equation}
        \gamma = \left(\frac{A}{\sigma_{\rm p}}\right)^2\frac{N}{2},
        \label{gamma}
        \end{equation}

        where A is the fraction of the amplitude of polarization variability:
        
    \begin{equation}
        A = \frac{|q_{\rm max} - q_{\rm min}| + |u_{\rm max} - u_{\rm min}|}{4},
        \label{amplitude}
        \end{equation}

        and $\sigma_{\rm p}$ is a standard deviation that is determined from the scatter of the observed Stokes parameters around the best fit curves, $N$ is the number of observations, and $q_{\rm max}$, $q_{\rm min}$, $u_{\rm max}$, $u_{\rm min}$ are the maximum and minimum values of the fitted Stokes parameters $q$ and $u$. The values of $\gamma$ we derived for the $B$, $V$, and $R$ passbands are 380, 345, and 320, respectively. 
        
        In order to estimate the bias and 1$\sigma$ confidence intervals for $i$ and $\Omega$, we used the plots given in (\citet{WD}, Figs. 4 and 6 therein). The resulting de-biased values of $i$ and $\Omega$ averaged over $B$, $V$, and $R$ passbands are $63^{\circ}+2^{\circ}/-3^{\circ}$ and $29^{\circ}(209^{\circ})\pm8^{\circ}$, respectively. Furthermore, we compared our derived value of $i$ with the values given by other sources. \citet{SHNDR} obtained the value of $51.7^{\circ}$ from the analysis of photometric data, \citet{RK} used polarimetry to estimate the value of $i$ from $57^{\circ}$ to $60^{\circ}$; \citet{PFR} also used polarization to estimate an inclination $\sim 55^{\circ}$ and \citet{GW} used spectroscopic observations to derive the value of $i = 61.1^{\circ}\pm3^{\circ}$. The modeling of \citet{PRGN}  gave a value $i \sim 65.7^{\circ}$. Orbital inclination derived from our polarimetry falls somewhere in between of those derived by \citet{GW} and \citet{PRGN}, and it is in good agreement with both.

    \section{Conclusions}\label{sec:conclusions}
        
    Our new polarimetric study of O-type binary star system AO~Cas has re-affirmed the existence of an intrinsic linear polarization, which is variable with the phase of binary orbital period. A Lomb-Scargle algorithm clearly detected an orbital period signal in our high-precision polarimetric data with no other frequency present. The amplitude of this variability is $\sim$0.2\% and now we are able to ascertain that most of the polarization observed from AO~Cas is of an interstellar nature. The small value of intrinsic polarization allowed us to rule out the presence of large circumstellar scattering envelope, as previously suggested by \cite{PFR}. Our new results indicate that light scattering material in AO~Cas is confined into a relatively small region and doesn't form an extended envelope around the system.
        
    From a comparison of our variability curve in the $B$ passband with that obtained by \cite{RK}, we have found that they are remarkably similar in amplitude and shape. This gives strong arguments in favor of a stable distribution of light scattering material in the AO~Cas binary system. We have not found evidence of seasonal changes in the intrinsic polarization in AO~Cas. Our new data have confirmed the presence of small (but not negligible) first-harmonics variations in Stokes $u$ parameter, pointing to some asymmetry in the distribution of scattering material in the binary system.  From the two-harmonics Fourier fit to the observed variations in the Stokes parameters, the estimates for the inclination, $i = 63^{\circ}+2^{\circ}/-3^{\circ}$, and orientation of the orbit, $\Omega = 29^{\circ}(209^{\circ})\pm8^{\circ}$. The direction of motion on this orbit, as seen on the sky, is clockwise. The inclination of the orbit derived from polarimetry is in a good agreement with the latest determinations made with other methods.
    
    We intend to continue our polarimetric studies of O-type binary star systems, originally initiated by \citet{Berd2016}, who discovered variable phase-locked polarization in HD 48099. We have observed another similar binary system, namely, DH Cephei. The initial data analysis shows phase-dependent polarization variability in DH Cep, just as in HD 48099 and AO~Cas. In the near future, we will present the results of our numerical scattering code modeling for AO~Cas and DH Cep systems. This modeling will certainly help in shedding more light on the distribution of light-scattering material in O-type binaries.

    \begin{acknowledgements}
      This work was supported by the ERC Advanced Grant Hot-
    Mol ERC-2011-AdG-291659 (www.hotmol.eu). Dipol-2 was built in the cooperation 
    between the University of Turku, Finland, and the Kiepenheuer Institut
    f\"{u}r Sonnenphysik, Germany, with the support by the Leibniz Association grant
    SAW-2011-KIS-7. We are grateful to the Institute for Astronomy, University of
    Hawaii for the observing time allocated for us on the T60 telescope at the Haleakal\={a} Observatory. All raw data and calibrations are available on request from the authors.
    \end{acknowledgements}

    \bibliography{allbib}
    \bibliographystyle{aa}

    \appendix
    \label{Appendix}
    \section{Log of polarimetric observations}

        \begin{table}[htp!]
        \caption{Log of polarimetric observations for AO~Cas.} 
        \label{table:log}
        \centering
        \renewcommand{\arraystretch}{1.0}                                                              \begin{tabular}[c]{l  c  c  c}
        \hline\hline 
        Date & MJD & $T_{\rm exp}~[\rm s]$ &  $N_{\rm obs}$ \\ \hline
        2020--10--26 & 59148.97 & 240 & 80 \\
        2020--10--31 & 59153.89 & 240 & 80 \\
        2020--11--01 & 59154.87 & 240 & 80 \\
        2020--11--03 & 59156.88 & 240 & 80 \\
        2020--11--04 & 59157.86 & 243 & 81 \\
        2020--11--05 & 59158.89 & 150 & 50 \\
        2020--11--08 & 59161.84 & 240 & 80 \\
        2020--11--09 & 59162.90 & 216 & 72 \\
        2020--11--10 & 59163.84 & 240 & 80 \\
        2020--11--13 & 59166.82 & 240 & 80 \\
        2020--11--14 & 59167.88 & 240 & 80 \\
        2020--11--16 & 59169.82 & 240 & 80 \\
        2020--11--17 & 59170.85 & 240 & 80 \\
        2020--11--19 & 59172.81 & 240 & 80 \\
        2020--11--20 & 59173.82 & 240 & 80 \\
        2020--11--26 & 59179.81 & 135 & 45 \\
        2020--11--27 & 59180.82 & 240 & 80 \\
        2020--11--30 & 59183.81 & 240 & 80 \\
        2020--12--01 & 59184.78 & 240 & 80 \\
        2020--12--02 & 59185.78 & 144 & 48 \\
        2020--12--25 & 59208.83 & 240 & 80 \\
        2020--12--26 & 59209.80 & 240 & 80 \\
        2020--12--28 & 59211.80 & 240 & 80 \\
        2020--12--29 & 59212.81 & 240 & 80 \\
        2021--01--02 & 59216.80 & 240 & 80 \\
        2021--01--03 & 59217.79 & 240 & 80 \\
        2021--01--05 & 59219.80 & 240 & 80 \\
        2021--01--06 & 59220.80 & 240 & 80 \\
        2021--01--07 & 59221.80 & 240 & 80 \\
        2021--01--11 & 59225.81 & 240 & 80 \\
        2021--01--12 & 59226.79 & 240 & 80 \\
        2021--01--13 & 59227.80 & 240 & 80 \\
        2021--01--14 & 59228.81 & 240 & 80 \\
        2021--01--15 & 59229.79 & 240 & 80 \\
        2021--01--16 & 59230.79 & 240 & 80 \\
        2021--01--28 & 59242.76 & 240 & 80 \\
        2021--01--29 & 59243.75 & 240 & 80 \\
        2021--01--31 & 59245.75 & 240 & 80 \\
        2021--02--01 & 59246.75 & 240 & 80 \\
        \hline
        \end{tabular}
        \end{table}

\end{document}